\documentclass[amssymb,aps,pre,showpacs,manuscript,showkeys]{revtex4}
\usepackage{mathrsfs}
\usepackage{amsmath}
\usepackage{amssymb}
\usepackage{graphicx}
\usepackage{esint}
\usepackage{mathrsfs}
\usepackage{amsthm}

\begin{document}

\title{Escape Process and Stochastic Resonance Under Noise Intensity Fluctuation}

\author{Yoshihiko Hasegawa}

\email{hasegawa@cb.k.u-tokyo.ac.jp}

\affiliation{Department of Biophysics and Biochemistry, Graduate School of Science,
The University of Tokyo, Tokyo 113-0033, Japan}

\author{Masanori Arita}

\affiliation{Department of Biophysics and Biochemistry, Graduate School of Science,
The University of Tokyo, Tokyo 113-0033, Japan}

\affiliation{Institute for Advanced Biosciences, Keio University, Yamagata 997-0035,
Japan}

\date{July 28, 2011}
\begin{abstract}
We study the effects of noise-intensity fluctuations on the stationary
and dynamical properties of an overdamped Langevin model with a bistable
potential and external periodical driving force. We calculated the
stationary distributions, mean-first passage time (MFPT) and the spectral
amplification factor using a complete set expansion (CSE) technique.
We found resonant activation (RA) and stochastic resonance (SR) phenomena
in the system under investigation. Moreover, the strength of RA and
SR phenomena exhibit non-monotonic behavior and their trade-off relation
as a function of the squared variation coefficient of the noise-intensity
process. The reliability of CSE is verified with Monte Carlo simulations.
\end{abstract}



\maketitle

\section{Introduction\label{sec:introduction}}

Langevin models have become increasingly important in modeling systems
subject to fluctuations. These models have a wide range of applications
in physics, chemistry, electronics, biology, and financial market
analysis. In many applications, fluctuations are modeled in terms
of white noise, which has a delta function correlation with constant
noise intensity. In general, fluctuations are space-time dependent
phenomena; hence, the noise intensity fluctuates temporally and/or
spatially. Nevertheless, white noise has been be used to model fluctuations
because at the typical level of physical description, variations in
noise intensity can be ignored. However, if the variation in the noise
intensity fluctuations is large and if it occurs in time scales comparable
to the physical description of interest, the effects of such fluctuations
have to be taken into account. Noise intensity fluctuations due to
environmental variations are particularly important in biological
applications. For instance, the stochasticity of a gene expression
mechanism is derived from intrinsic (discreteness of particle number)
and extrinsic (noise sources external to the system) fluctuations.
Because extrinsic fluctuations are subject to biological rhythms with
different time scales \cite{Novak:2008:BiochemOsc}, their noise intensity
varies temporally.

In financial market analysis, stochastic volatility models (e.g.,
the Hull \& White model and Heston model) incorporate temporal noise
intensity fluctuations \cite{Hull:1987:SV,Heston:1993:Volatility,Dragulescu:2002:Heston,Andersson:2003:SV}.
The stochastic volatility models assume that noise variance is governed
by stochastic processes. In physics, superstatistics \cite{Wilk:2000:NEXTParam,Beck:2001:DynamicalNEXT,Beck:2003:Superstatistics,Beck:2006:SS_Brownian,Beck:2009:RecentSS,Beck:2010:GStatMech}
take spatial and/or temporal environmental fluctuations into account.
Superstatistics has been applied to stochastic processes, and it has
introduced noise intensity fluctuations \cite{Beck:2001:DynamicalNEXT,Beck:2006:SS_Brownian,Jizba:2008:SuposPD,Queiros:2008:SSMultiplicative,Hasegawa:2010:qExpBistable,Rodriguez:2007:SS_Brownian,Straeten:2010:SkewedSS},
by calculating stationary distributions in a Bayesian manner. A previous
study \cite{Queiros:2005:VolatilityNEXT} indicates the similarity
between distributions of a stochastic volatility model and Tsallis
statistics, which has the same stationary distribution ($q$-Gaussian
distribution) as superstatistics in specific cases.

Most discussions on stochastic volatility models are limited to linear
drift terms; hence, the application of such models to physical, chemical,
or biological systems, accompanied by nontrivial drift terms and multiplicative
noise, is nontrivial. In our previous paper \cite{Hasegawa:2010:SIN},
we proposed an approximation scheme that can be applied to general
drift terms. We considered Langevin equations where the white Gaussian
noise intensity is governed by the Ornstein--Uhlenbeck process: 
\begin{eqnarray}
\frac{dx}{dt} & = & f(x)+s\xi_{x}(t),\label{eq:main_Langevin}\\
\frac{ds}{dt} & = & -\gamma(s-\alpha)+\sqrt{\gamma}\xi_{s}(t),\label{eq:SIN_Langevin}
\end{eqnarray}
 where $f(x)$ is a drift term ($f(x)=-\partial_{x}U(x)$, where $U(x)$
is a potential), $\gamma$ is the relaxation rate, and $\xi_{x}(t)$,
$\xi_{s}(t)$ denote white Gaussian noise with the correlation {[}$\left\langle \xi_{x}(t)\xi_{x}(t^{\prime})\right\rangle =2D_{x}\delta(t-t^{\prime})$
and $\left\langle \xi_{s}(t)\xi_{s}(t^{\prime})\right\rangle =2D_{s}\delta(t-t^{\prime})${]}.
In the present paper, we call the term $s\xi_{x}(t)$ the stochastic
intensity noise (SIN) because the noise intensity is governed by a
stochastic process. In Ref.~\cite{Hasegawa:2010:SIN}, we obtained
a time evolution equation using adiabatic elimination with an eigenfunction
expansion \cite{Kaneko:1976:AdiabaticElim}. Although the previously
developed method \cite{Hasegawa:2010:SIN} can be applied to nonlinear
drift terms, its application is limited to $\gamma\gg1$ {[}$\gamma$
is the relaxation rate in Eq.~(\ref{eq:SIN_Langevin}){]}. At the
same time, we showed that the time evolution equation of $P(x;t)$
is a higher order Fokker--Planck equation (FPE) having derivatives
of orders higher than two \cite{Hasegawa:2010:SIN}. Analytic calculations
of dynamical quantities such as mean-first passage time (MFPT) and
stochastic resonance (SR) are mainly developed for one-variable FPEs;
hence, their use in higher-order FPEs is nontrivial. Accordingly,
in this paper, we investigate the dynamical properties of the coupled
equations (\ref{eq:main_Langevin}) and (\ref{eq:SIN_Langevin}),
expanding functions of interest (stationary distributions and eigenfunctions)
in terms of an orthonormal complete set. This technique is extensively
used to solve FPEs numerically (e.g., the matrix continued fraction
method. For details, please see Ref.~\cite{Risken:1989:FPEBook}
and the references therein). Complete set expansion (CSE) can be applied
to polynomial drift terms, and it can, in principle, solve for the
entire range of $\gamma$; on the other hand, the adiabatic elimination
based method is limited to $\gamma\gg1$ \cite{Hasegawa:2010:SIN}.

In the present paper, we investigate MFPT and SR with a bistable potential
{[}see Eq.~(\ref{eq:quartic_bistable}){]}. As stated above, SIN
is particularly important in biological mechanisms. In a zeroth-order
approximation, many important biological mechanisms, such as neuron
and gene expression, can be modeled with a bistable potential. MFPT
and SR have also been extensively investigated in such biological
mechanisms. In the calculation of MFPT, we show that MFPT, as a function
of $\gamma$, has a minimum around $\gamma\simeq1$, which is equivalent
to resonant activation (RA) \cite{Doering:1992:ResonantActivation,Zurcher:1993:ResoAct,Marchi:1996:ResoAct,Boguna:1998:RAproperties,Mantegna:2000:ExpRA,Fiasconaro:2011:AsymRA}.
Furthermore, by changing $\rho$ (the squared variation coefficient
of noise intensity fluctuations {[}see Eq.~(\ref{eq:rho_def}){]}),
MFPT also has a minimum around $\rho\simeq1$. In the calculation
of SR, we show that the SR effect is smaller for smaller $\gamma$,
which indicates that the SR effect is maximized under white noise.
In addition, the spectral amplification factor, as a function of $\rho$,
has a minimum around $\rho\simeq1$. These results show that the strength
of RA and SR effects cannot be maximized simultaneously. All the calculations
are performed using CSE, whose reliability is evaluated via Monte
Carlo (MC) simulations.

The remainder of this paper is organized as follows. In Sec.~\ref{sec:model},
we describe the model adopted in this study. In Sec.~\ref{sec:stat_dist},
stationary distributions are calculated using CSE. In Sec.~\ref{sec:MFPT},
we calculate MFPT, which is approximated by the smallest non-vanishing
eigenvalue. In Sec.~\ref{sub:SR}, we investigate the spectral amplification
factor of SR by using the linear response approximation. In Sec.~\ref{sec:discussion},
we discuss the effects of noise intensity fluctuations on RA and SR.
Finally, in Sec.~\ref{sec:remarks}, we conclude the paper.

\section{The Model\label{sec:model}}

We consider the Langevin equations given by Eqs.~(\ref{eq:main_Langevin})
and (\ref{eq:SIN_Langevin}) with the bistable potential 
\begin{equation}
U(x)=\frac{x^{4}}{4}-\frac{x^{2}}{2},\label{eq:quartic_bistable}
\end{equation}
 \emph{i.e.,} $f(x)=x-x^{3}$. In this paper, we investigate the $\gamma>0$
case for Eq.~(\ref{eq:SIN_Langevin}) because $s(t)$ is constant
($s(t)=s(0)$) for $\gamma=0$, and the resulting SIN is equivalent
to conventional white Gaussian noise.

By interpreting Eqs.~(\ref{eq:main_Langevin}) and (\ref{eq:SIN_Langevin})
in the Stratonovich sense, a probability density function $P(x,s;t)$
of $(x,s)$ at time $t$ is governed by the FPE: 
\begin{equation}
\frac{\partial}{\partial t}P(x,s;t)=\mathscr{L}_{0}P(x,s;t),\label{eq:basic_FPE}
\end{equation}
 where $\mathscr{L}_{0}$ is an FPE operator defined as 
\begin{equation}
\mathscr{L}_{0}=\mathscr{L}_{x}+\gamma\mathscr{L}_{s},\label{eq:FPE_add}
\end{equation}
 with 
\begin{equation}
\mathscr{L}_{x}=-\frac{\partial}{\partial x}f(x)+s^{2}D_{x}\frac{\partial^{2}}{\partial x^{2}},\label{eq:FPE_Lx}
\end{equation}
 
\begin{equation}
\mathscr{L}_{s}=\frac{\partial}{\partial s}(s-\alpha)+D_{s}\frac{\partial^{2}}{\partial s^{2}}.\label{eq:FPE_Ls}
\end{equation}
 For the asymptotic case $\gamma\rightarrow\infty$, we used the adiabatic
elimination technique to obtain the FPE operator \cite{Hasegawa:2010:SIN}
\begin{equation}
\mathscr{L}_{0}=-\frac{\partial}{\partial x}f(x)+Q\frac{\partial^{2}}{\partial x^{2}}\hspace{1em}(\mathrm{for}\,\,\,\,\gamma\rightarrow\infty),\label{eq:gamma_inf_op}
\end{equation}
 where $Q$ is the effective noise intensity given by 
\begin{equation}
Q=D_{x}(D_{s}+\alpha^{2}).\label{eq:Q_def}
\end{equation}
 Equation~(\ref{eq:Q_def}) is in agreement with the noise intensity
of the correlation function, \emph{i.e.,} $\left\langle s(t)\xi_{x}(t)s(t^{\prime})\xi_{x}(t^{\prime})\right\rangle =2Q\delta(t-t^{\prime})$
(see the Appendix).

\begin{figure}
\begin{centering}
\includegraphics[width=15cm]{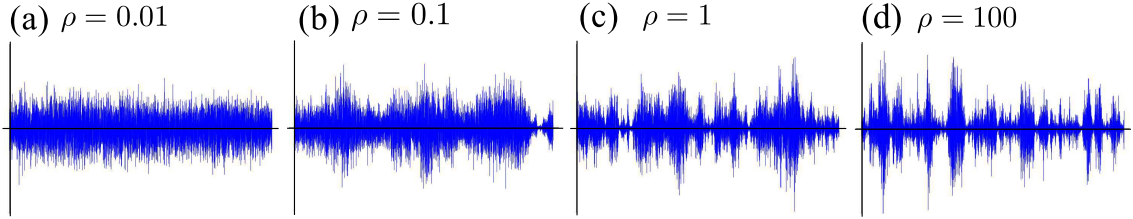} 
\par\end{centering}

\caption{(Color online) Trajectories of SIN for four parameter values of $\rho$
(squared variation coefficient): (a) $\rho=0.01$, (b) $\rho=0.1$,
(c) $\rho=1$, and (d) $\rho=100$. We varied $\rho$ while keeping
the effective intensity $Q$ constant. \label{fig:path_rho}}
\end{figure}

From Eq.~(\ref{eq:FPE_Ls}), the stationary distribution $P_{st}(s)$
of the intensity-modulating term $s$ is given by 
\begin{equation}
P_{st}(s)=\frac{1}{\sqrt{2\pi D_{s}}}\exp\left\{ -\frac{1}{2D_{s}}(s-\alpha)^{2}\right\} .\label{eq:statdist_s}
\end{equation}
 Here, we introduce the squared variation coefficient of the noise
intensity fluctuation for later use. The squared variation coefficient
$\rho$ is defined as 
\begin{equation}
\rho=\frac{D_{s}}{\alpha^{2}},\label{eq:rho_def}
\end{equation}
 where $\rho$ denotes the squared ratio between the standard deviation
and mean of Eq.~(\ref{eq:statdist_s}), similar to the Fano factor.
Figure \ref{fig:path_rho} shows some trajectories of SIN with (a)
$\rho=0.01$, (b) $\rho=0.1$, (c) $\rho=1$, and (d) $\rho=100$.
These trajectories have the same effective noise intensity $Q$. As
$\rho\rightarrow0$, SIN reduces to white Gaussian noise with noise
intensity $Q=D_{x}\alpha^{2}$.

In the present paper, the FPE of Eq.~(\ref{eq:basic_FPE}) is solved
using CSE and MC. MC is performed by adopting the Euler forward method
with time resolution $\Delta t=10^{-4}$ (for details of the method,
see Ref.~\cite{Risken:1989:FPEBook}).

\section{Stationary Distributions\label{sec:stat_dist}}

We calculate the stationary distributions of the coupled Langevin
equations (\ref{eq:main_Langevin}) and (\ref{eq:SIN_Langevin}),
which have been discussed previously \cite{Hasegawa:2010:SIN} for
$\gamma\gg1$. The method adopted in this paper is different from
the previous one \cite{Hasegawa:2010:SIN} in the range of $\gamma$
(previously \cite{Hasegawa:2010:SIN}, it was limited to $\gamma\gg1$).
In the following, we first investigate the effects of noise intensity
fluctuations on the stationary distributions. Then, the calculations
of the stationary distributions are used for the spectral amplification
factor in SR (Sec.~\ref{sub:SR}).

The stationary distribution $P_{0}(x,s)$ of $(x,s)$ has to satisfy
the differential equation: 
\begin{equation}
\mathscr{L}_{0}P_{0}(x,s)=0,\label{eq:stat_FPE_op}
\end{equation}
 where $\mathscr{L}_{0}$ is an FPE operator defined in Eq.~(\ref{eq:basic_FPE}).
In order to solve Eq.~(\ref{eq:stat_FPE_op}), we employ CSE, which
expands $P_{0}(x,s)$ in terms of an orthonormal complete set. This
technique is extensively used in stochastic processes (e.g., the matrix
continued fraction method \cite{Risken:1989:FPEBook}). CSE can handle
systems with polynomial drift terms and it can, in principle, handle
the entire range of $\gamma$. However, in practical calculations,
we are restricted to $\gamma\ge0.3$ because of numerical instability.
Considering the symmetry $x\rightarrow-x$ in $\mathscr{L}_{0}$ and
the relation $\varphi_{2k}(-x)=\varphi_{2k}(x)$, the stationary distribution
$P_{0}(x,s)$ admits the even parity expansion: 
\begin{equation}
P_{0}(x,s)=\varphi_{0}(x)\psi_{0}(s)\sum_{k=0}^{K}\sum_{\ell=0}^{L}C_{k,\ell}\varphi_{2k}(x)\psi_{\ell}(s),\label{eq:ortho_expansion}
\end{equation}
 with 
\begin{equation}
\varphi_{k}(x)=\sqrt{\frac{\zeta}{2^{k}k!\sqrt{\pi}}}H_{k}(\zeta x)\exp\left(-\frac{1}{2}\zeta^{2}x^{2}\right),\label{eq:complete_set_x}
\end{equation}
 
\begin{equation}
\psi_{\ell}(s)=\left(\frac{1}{2\pi D_{s}}\right)^{1/4}\sqrt{\frac{1}{2^{\ell}\ell!}}H_{\ell}\left(\eta\right)\exp\left(-\frac{1}{2}\eta^{2}\right).\label{eq:complete_set_s}
\end{equation}
 Here, $C_{k,\ell}$ are expansion coefficients, $\eta=\sqrt{1/(2D_{s})}(s-\alpha)$,
$H_{n}(z)$ is the $n$th Hermite polynomial, and $\zeta$ is a (positive)
scaling parameter that affects the convergence of CSE. $K$ and $L$
are truncation numbers which provide the precision of the obtained
solutions. The orthonormality and complete relations read 
\begin{equation}
\int dx\,\varphi_{k'}(x)\varphi_{k}(x)=\delta_{k',k},\hspace{1em}\int ds\,\psi_{\ell'}(s)\psi_{\ell}(s)=\delta_{\ell',\ell},\label{eq:ortho_complete_relation}
\end{equation}
 where $\delta_{k',k}$ is Kronecker's delta function. The term $\psi_{0}(s)\psi_{\ell}(s)$
forms eigenfunctions of $\mathscr{L}_{s}$ {[}Eq.~(\ref{eq:FPE_Ls}){]},
\emph{i.e.,} 
\begin{equation}
\mathscr{L}_{s}\left[\psi_{0}(s)\psi_{\ell}(s)\right]=-\ell\psi_{0}(s)\psi_{\ell}(s).\label{eq:eigen_Ls}
\end{equation}
 After multiplying $\varphi_{2k'}(x)\psi_{\ell'}(s)/(\varphi_{0}(x)\psi_{0}(s))$
by Eq.~(\ref{eq:stat_FPE_op}) and integrating with respect to $x$
and $s$, we obtain the following linear algebraic equation: 
\begin{eqnarray}
0 & = & C_{k,\ell}\left(2k-\frac{6k^{2}}{\zeta^{2}}-\gamma\ell\right)\nonumber \\
 &  & +C_{k-1,\ell}\sqrt{2k(2k-1)}\left[1-\frac{3}{2\zeta^{2}}(2k-1)+2\zeta^{2}D_{x}\left\{ \alpha^{2}+2D_{s}\left(\ell+\frac{1}{2}\right)\right\} \right]\nonumber \\
 &  & -C_{k+1,\ell}\frac{k}{\zeta^{2}}\sqrt{(2k+1)(2k+2)}-C_{k-2,\ell}\frac{1}{2\zeta^{2}}\sqrt{2k(2k-1)(2k-2)(2k-3)}\nonumber \\
 &  & +2C_{k-1,\ell+2}\zeta^{2}D_{x}D_{s}\sqrt{2k(2k-1)(\ell+2)(\ell+1)}+2C_{k-1,\ell-2}\zeta^{2}D_{x}D_{s}\sqrt{2k(2k-1)\ell(\ell-1)}\nonumber \\
 &  & +4C_{k-1,\ell-1}\zeta^{2}\alpha D_{x}\sqrt{2D_{s}k(2k-1)\ell}+4C_{k-1,\ell+1}\zeta^{2}\alpha D_{x}\sqrt{2D_{s}k(2k-1)(\ell+1)}.\label{eq:P0_linear}
\end{eqnarray}
 Because all coefficients vanish for $(k,\ell)=(0,0)$, $C_{0,0}$
can be determined by a normalization condition {[}$\int ds\int dx\, P_{0}(x,s)=1${]}
as $C_{0,0}=1$. The two-dimensional coefficients $C_{k,\ell}$ can
be cast in the form of one-dimensional coefficients $\mathcal{C}_{m}$
by the following one-to-one mapping \cite{Denisov:2009:ACmotorFPE}:
\begin{equation}
m=1+(1+L)k+\ell.\label{eq:C_mapping}
\end{equation}
 By using Eq.~(\ref{eq:C_mapping}), $C_{k,\ell}$ can be transformed
into $\mathcal{C}_{m}$ with $1\le m\le M$, where $M=(1+K)(1+L)$.
Eq.~(\ref{eq:P0_linear}) can be solved using general linear algebraic
solvers. CSE transforms the differential equations into linear algebraic
equations, which are easier to solve. From Eq.~(\ref{eq:gamma_inf_op}),
the stationary distribution $P_{0}(x)$ of $x$ in the asymptotic
case $\gamma\rightarrow\infty$ is given by 
\begin{equation}
P_{0}(x)=\int ds\, P_{0}(x,s)=\frac{1}{Z}\exp\left(-\frac{U(x)}{Q}\right)\hspace{1em}(\mathrm{for}\,\,\,\gamma\rightarrow\infty),\label{eq:statdist_inf}
\end{equation}
 where $Z$ is a normalizing constant.

\begin{figure}
\begin{centering}
\includegraphics[width=7cm]{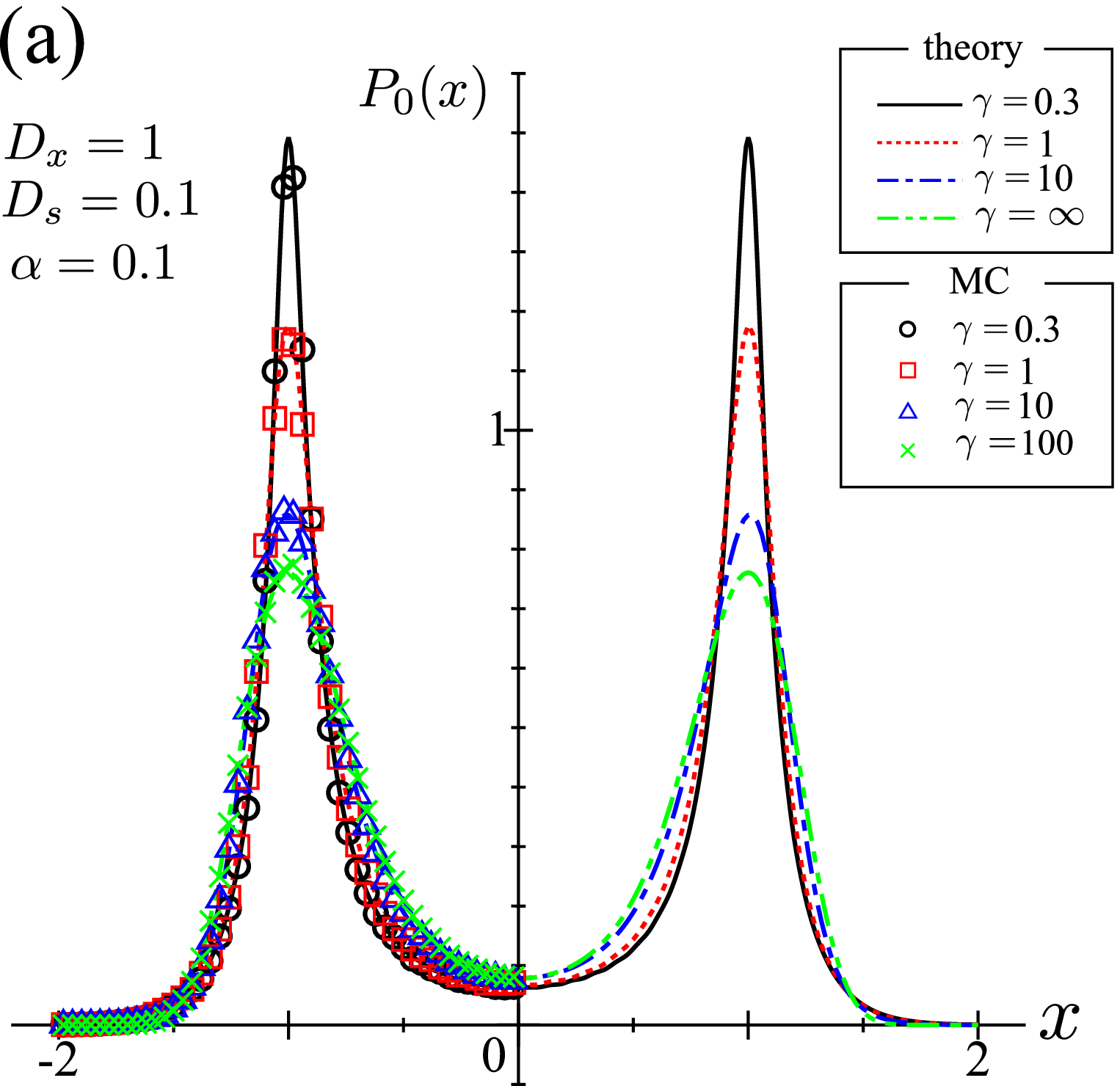}~~~\includegraphics[width=7cm]{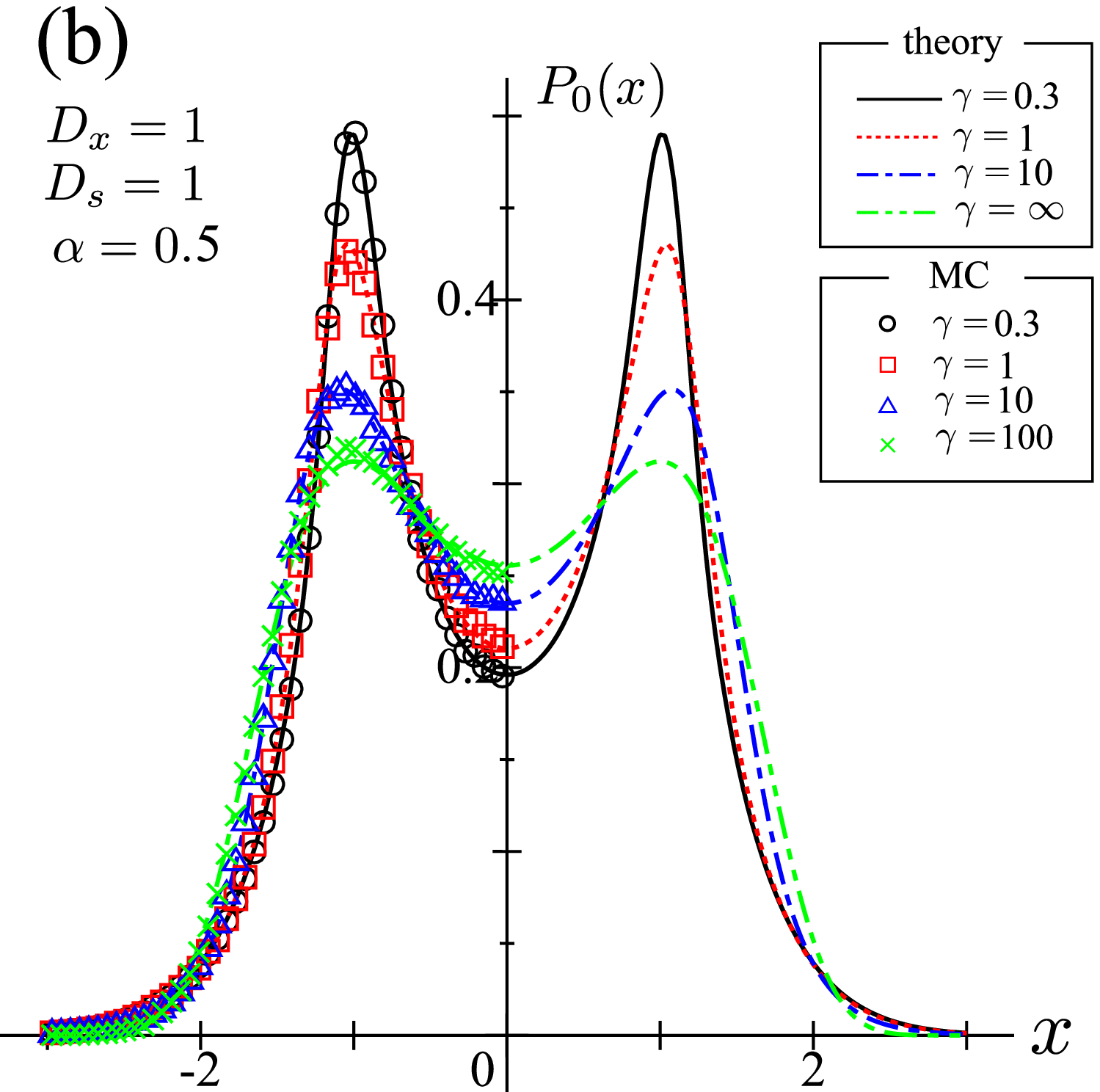} 
\par\end{centering}

\caption{(Color online) Stationary distributions for systems driven by additive
SIN. The lines and symbols represent distributions calculated using
CSE and MC methods, respectively. The parameters are (a) $D_{x}=1$,
$D_{s}=0.1$, $\alpha=0.1$ and (b) $D_{x}=1$, $D_{s}=1$, $\alpha=0.5$,
with $\gamma=0.3$ (solid lines and circles), $1$ (dotted lines and
squares), and $10$ (dot-dashed lines and triangles). Furthermore,
the dot-dot-dashed lines are given by Eq.~(\ref{eq:statdist_inf}),
which corresponds to $\gamma\rightarrow\infty$, and the crosses denote
MC results with $\gamma=100$. For viewability, MC data are plotted
only for $x<0$. \label{fig:statdist}}
\end{figure}

In calculating stationary distributions using CSE, we have to determine
$K$, $L$, and $\zeta$. We increase $K$ and $L$ until the stationary
distributions converge. Although larger values of $K$ and $L$ allow
better approximation, we find that using excessively large values
numerically gives rise to divergent distributions. Fig.~\ref{fig:statdist}
shows stationary distributions with different parameters: $D_{x}=1$,
$D_{s}=0.1$, and $\alpha=0.1$ (Fig.~\ref{fig:statdist}(a)); and
$D_{x}=1$, $D_{s}=1$, and $\alpha=0.5$ (Fig.~\ref{fig:statdist}(b)).
Figs.~\ref{fig:statdist}(a) and (b) show stationary distributions
calculated using CSE for four $\gamma$ values: $\gamma=0.3$ (solid
line), $\gamma=1$ (dotted line), $\gamma=10$ (dot-dashed line),
and $\gamma\rightarrow\infty$ (dot-dot-dashed line). Although the
CSE method is valid, in principle, for the entire range of $\gamma$,
it appears that small values of $\gamma$ give rise to numerical instability.
Consequently, the smallest value used in this paper is $\gamma=0.3$.
For $\gamma\rightarrow\infty$, we used the asymptotic expression
given by Eq.~(\ref{eq:statdist_inf}). The stationary distributions
of MC simulations were computed for four $\gamma$ values: $\gamma=0.3$
(circles), $\gamma=1$ (squares), $\gamma=10$ (triangles), and $\gamma=100$
(crosses). Total $10^{6}$ samples each were calculated for the empirical
probability densities. Higher peaks emerge at metastable sites for
smaller $\gamma$. The CSE stationary distribution of $\gamma\rightarrow\infty$
and the MC stationary distribution of $\gamma=100$ are very close,
which supports the result that a system driven by SIN reduces to one
driven by white Gaussian noise with effective noise intensity $Q$.

\section{Mean First Passage Time\label{sec:MFPT}}

In order to study the dynamical properties of systems driven by SIN,
we calculate MFPT. With regard to the stochastic volatility model,
an escape problem was investigated for the extended Heston volatility
model in a cubic potential using MC simulations \cite{Bonanno:2007:EscapeSV}.
Non-monotonic phenomena such as noise-enhanced stability (NES) \cite{Mantegna:1996:NES,Spagnolo:2008:NES_review,Dubkov:2004:NES,Fiasconaro:2009:NESinColoredNoise}
were reported for this model. Another study \cite{Iwaniszewski:2008:RAtemperature}
considered a Langevin system, where the temperature (\emph{i.e.,}
noise intensity) takes two values in a random dichotomatic manner,
indicating the occurrence of an RA \cite{Doering:1992:ResonantActivation,Zurcher:1993:ResoAct,Marchi:1996:ResoAct,Boguna:1998:RAproperties,Mantegna:2000:ExpRA,Fiasconaro:2011:AsymRA}
phenomenon.

\begin{figure}
\begin{centering}
\includegraphics[width=7cm]{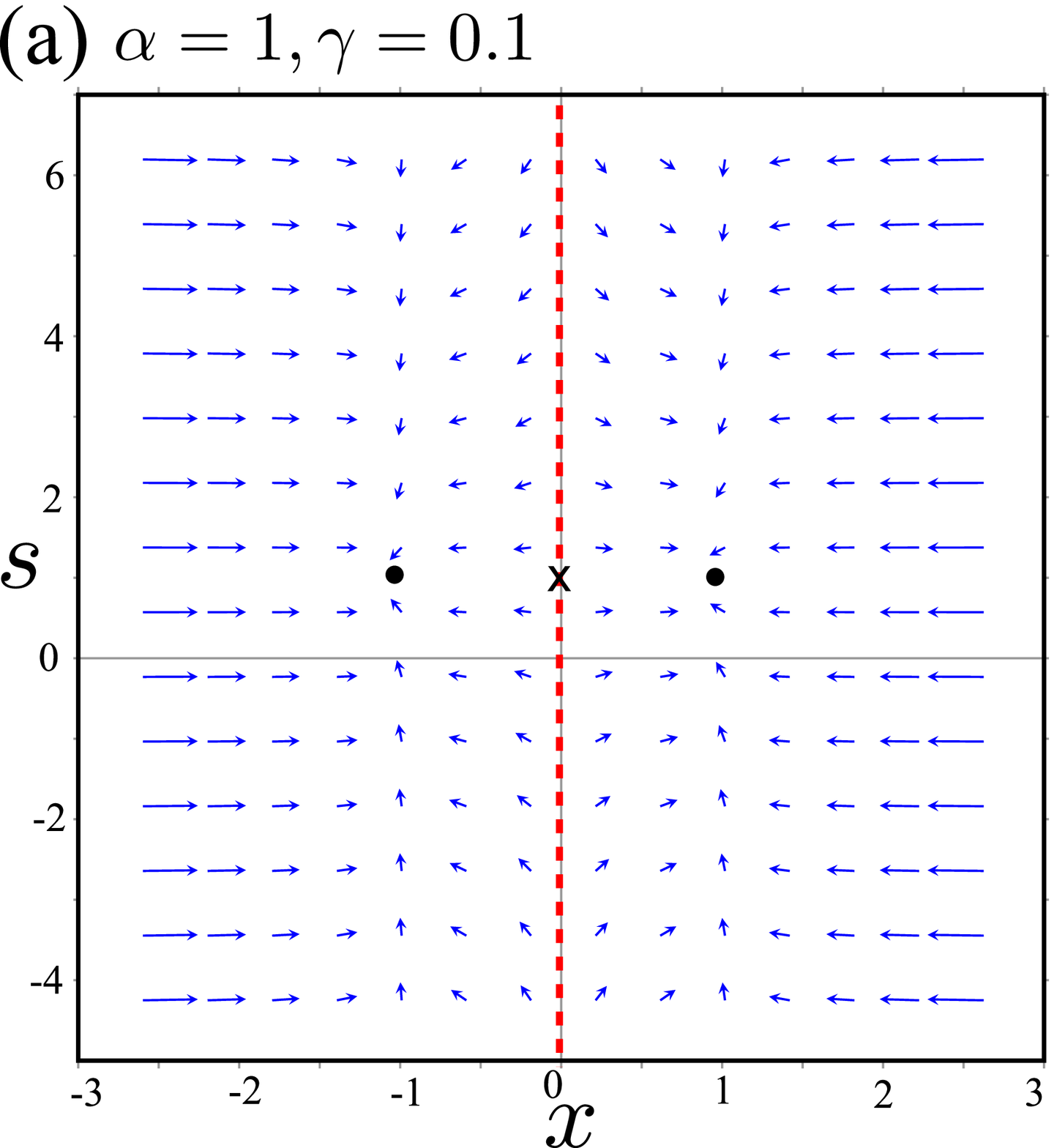}~~~~\includegraphics[width=7cm]{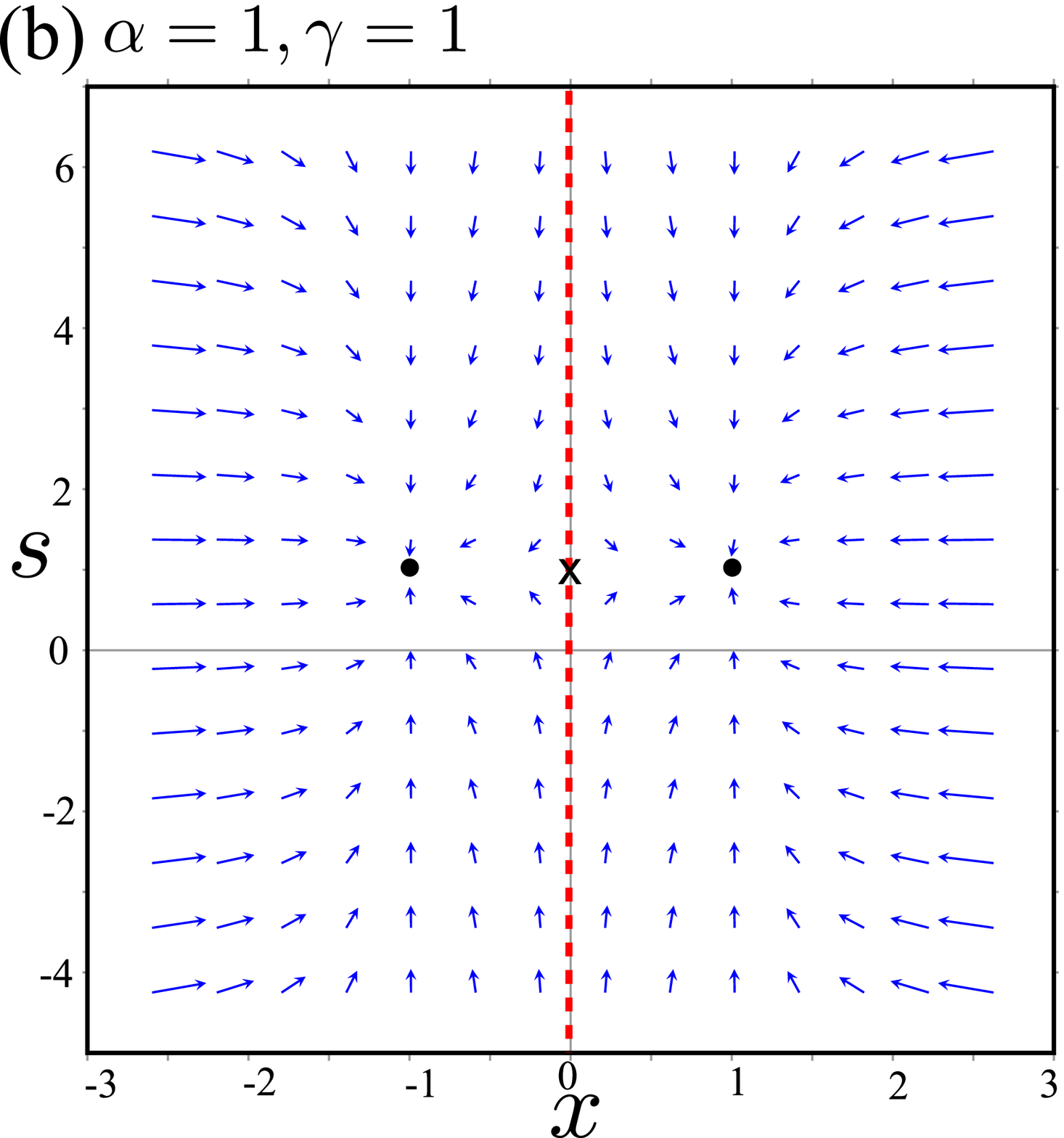} 
\par\end{centering}

\begin{centering}
\includegraphics[width=7cm]{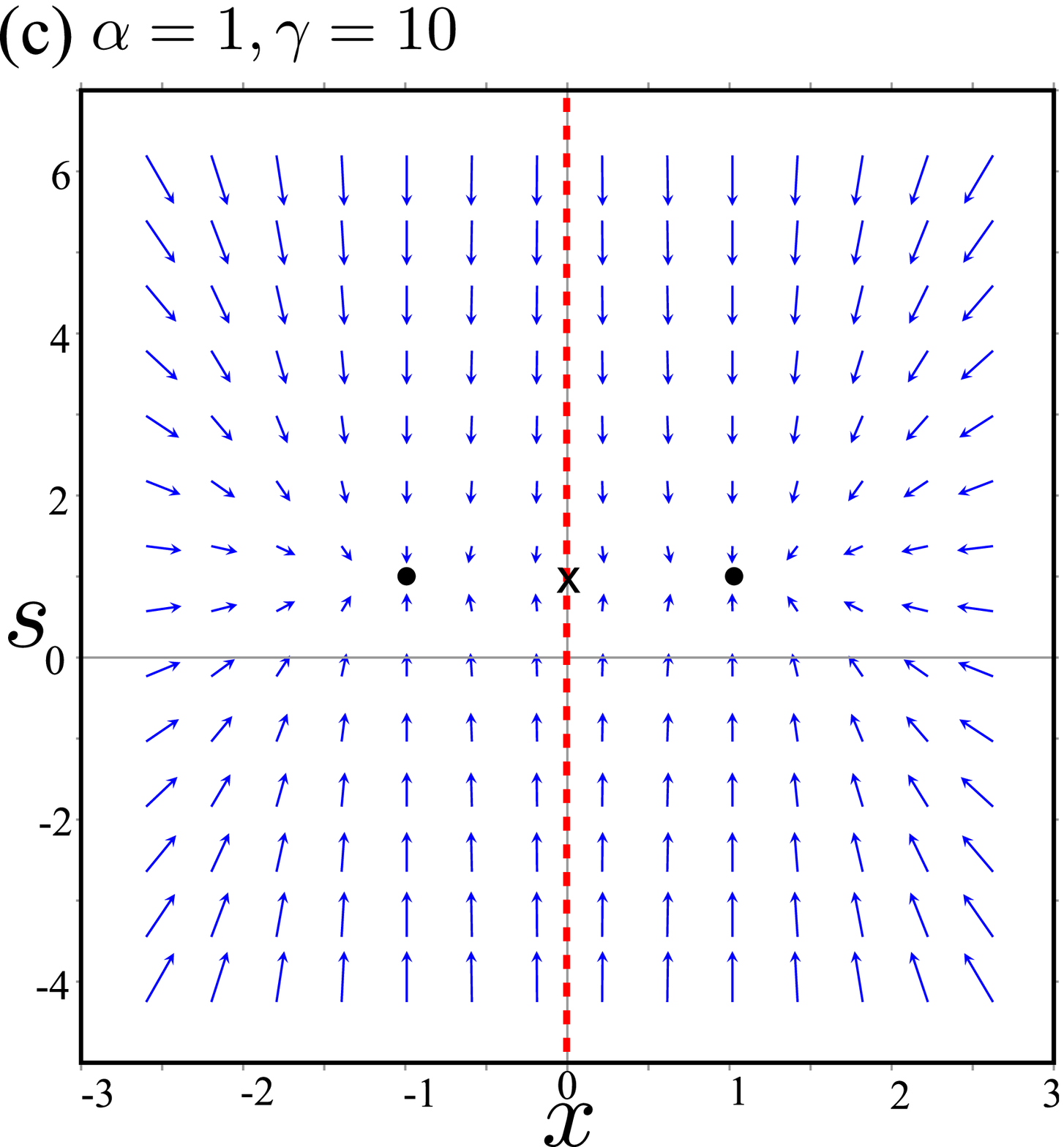} 
\par\end{centering}

\caption{(Color online) Vector field plots of Eq.~(\ref{eq:no_fluc}) with
$\alpha=1$ and (a) $\gamma=0.1$, (b) $\gamma=1$, and (c) $\gamma=10$.
There are three fixed points at $(\pm1,\alpha)$ (stables points)
and $(0,\alpha)$ (a saddle point), which are denoted by filled circles
and crosses, respectively. The dotted line is a separatix, which separates
the two basins. \label{fig:flow}}
\end{figure}

First, we investigate two basins of attractors and a separatix that
separates them in $(x,s)$ space. Without fluctuations, the deterministic
dynamics of Eqs.~(\ref{eq:main_Langevin}) and (\ref{eq:SIN_Langevin})
are given by 
\begin{equation}
\frac{dx}{dt}=f(x),\hspace{1em}\frac{ds}{dt}=-\gamma(s-\alpha).\label{eq:no_fluc}
\end{equation}
 Considering the quartic bistable potential $f(x)=x-x^{3}$, Eq.~(\ref{eq:no_fluc})
has three fixed points: $(\pm1,\alpha)$ (stable points) and $(0,\alpha)$
(a saddle point). Deterministic trajectories of Eq.~(\ref{eq:no_fluc})
are given by \cite{Hanggi:1989:EscapeCorNoise} 
\begin{equation}
\frac{ds}{dx}=-\frac{\gamma(s-\alpha)}{x-x^{3}}.\label{eq:flow_DE}
\end{equation}
 Specific trajectories, as a function of $x$, are obtained by solving
Eq.~(\ref{eq:flow_DE}): 
\begin{equation}
s(x)=\alpha+W|x|^{-\gamma}|x^{2}-1|^{\gamma/2},\label{eq:flow_line}
\end{equation}
 where $W$ is an integral constant. Figure~\ref{fig:flow} shows
vector field plots of Eq.~(\ref{eq:no_fluc}) for three $\gamma$
cases: (a) $\gamma=0.1$, (b) $\gamma=1$, and (c) $\gamma=10$. In
Fig.~\ref{fig:flow}, the dotted line represents the separatix. We
see that the separatix is $x=0$ regardless of $\gamma$, which is
not the case for colored-noise-driven systems (the separatix depends
on the time-correlation of colored noise).

Let $\tau_{s}$ be MFPT to the separatix ($x=0$). For sufficiently
low noise intensity, MFPT $\tau_{s}$ can be well approximated by
an eigenvalue \cite{Vollmer:1983:EigenvalueEscape,Jung:1993:PeriodicSystem,Jung:1989:ThermalActivation,Bartussek:1995:SurmountWell}:
\begin{equation}
\tau_{s}=\frac{1}{2r}=\frac{1}{\lambda_{1}},\label{eq:MFPT_eigenvalue}
\end{equation}
 where $r$ is the escape rate and $\lambda_{1}$ is the smallest
non-vanishing eigenvalue of the FPE operator $\mathscr{L}_{0}$ {[}Eq.~(\ref{eq:FPE_add}){]}.
Equation~(\ref{eq:MFPT_eigenvalue}) gives a reliable approximation
when the noise intensity is sufficiently small and $\lambda_{1}$
is well separated from the remaining eigenvalues {[}$\lambda_{n}$
($n\ge2$){]}. The eigenvalue problem is represented by the equation
\begin{equation}
\mathscr{L}_{0}\phi(x,s)=-\lambda\phi(x,s),\label{eq:eigen_relation}
\end{equation}
 where $\lambda$ and $\phi(x,s)$ are eigenvalues and eigenfunctions,
respectively. To calculate the eigenvalues, we employ CSE as in the
case of stationary distributions. According to the symmetry $x\rightarrow-x$
in $\mathscr{L}_{0}$, the eigenfunctions $\phi(x,s)$ have even {[}$\phi^{\mathrm{e}}(-x,s)=\phi^{\mathrm{e}}(x,s)${]}
or odd {[}$\phi^{\mathrm{o}}(-x,s)=-\phi^{\mathrm{o}}(x,s)${]} parity
symmetry. The even case expansion is identical to Eq.~(\ref{eq:ortho_expansion}),
and the odd case admits the following expansion: 
\begin{equation}
\phi^{\mathrm{o}}(x,s)=\varphi_{0}(x)\psi_{0}(s)\sum_{k=0}^{K}\sum_{\ell=0}^{L}C_{k,\ell}\varphi_{2k+1}(x)\psi_{\ell}(s).\label{eq:ev_odd_expansion}
\end{equation}
In the same procedure as that for stationary distributions, the even
and odd cases of Eq.~(\ref{eq:eigen_relation}) can be reduced to
linear algebraic equations. By using CSE, Eq.~(\ref{eq:eigen_relation})
for the odd case is calculated as 
\begin{eqnarray}
 &  & C_{k,\ell}\left\{ 2k+1-\frac{3}{2\zeta^{2}}(2k+1)^{2}-\gamma\ell\right\} -C_{k+1,\ell}\frac{2k+1}{2\zeta^{2}}\sqrt{(2k+2)(2k+3)}\nonumber \\
 &  & +C_{k-1,\ell}\sqrt{2k(2k+1)}\left[1-\frac{3k}{\zeta^{2}}+2\zeta^{2}D_{x}\left\{ \alpha^{2}+2D_{s}\left(\ell+\frac{1}{2}\right)\right\} \right]\nonumber \\
 &  & -C_{k-2,\ell}\frac{1}{2\zeta^{2}}\sqrt{(2k+1)2k(2k-1)(2k-2)}+2C_{k-1,\ell+2}\zeta^{2}D_{x}D_{s}\sqrt{2k(2k+1)(\ell+1)(\ell+2)}\nonumber \\
 &  & +2C_{k-1,\ell-2}\zeta^{2}D_{x}D_{s}\sqrt{2k(2k+1)\ell(\ell-1)}+4C_{k-1,\ell-1}\zeta^{2}D_{x}\alpha\sqrt{2D_{s}k(2k+1)\ell}\nonumber \\
 &  & +4C_{k-1,\ell+1}\zeta^{2}D_{x}\alpha\sqrt{2D_{s}k(2k+1)(\ell+1)}.\nonumber \\
 &  & =-\lambda C_{k,\ell}.\label{eq:linear_eq_eigen}
\end{eqnarray}
 Equation~(\ref{eq:eigen_relation}) is now transformed into a linear
algebraic eigenvalue problem, which can be solved with general linear
algebraic eigenvalue solvers.

\begin{figure}
\begin{centering}
\includegraphics[width=8cm]{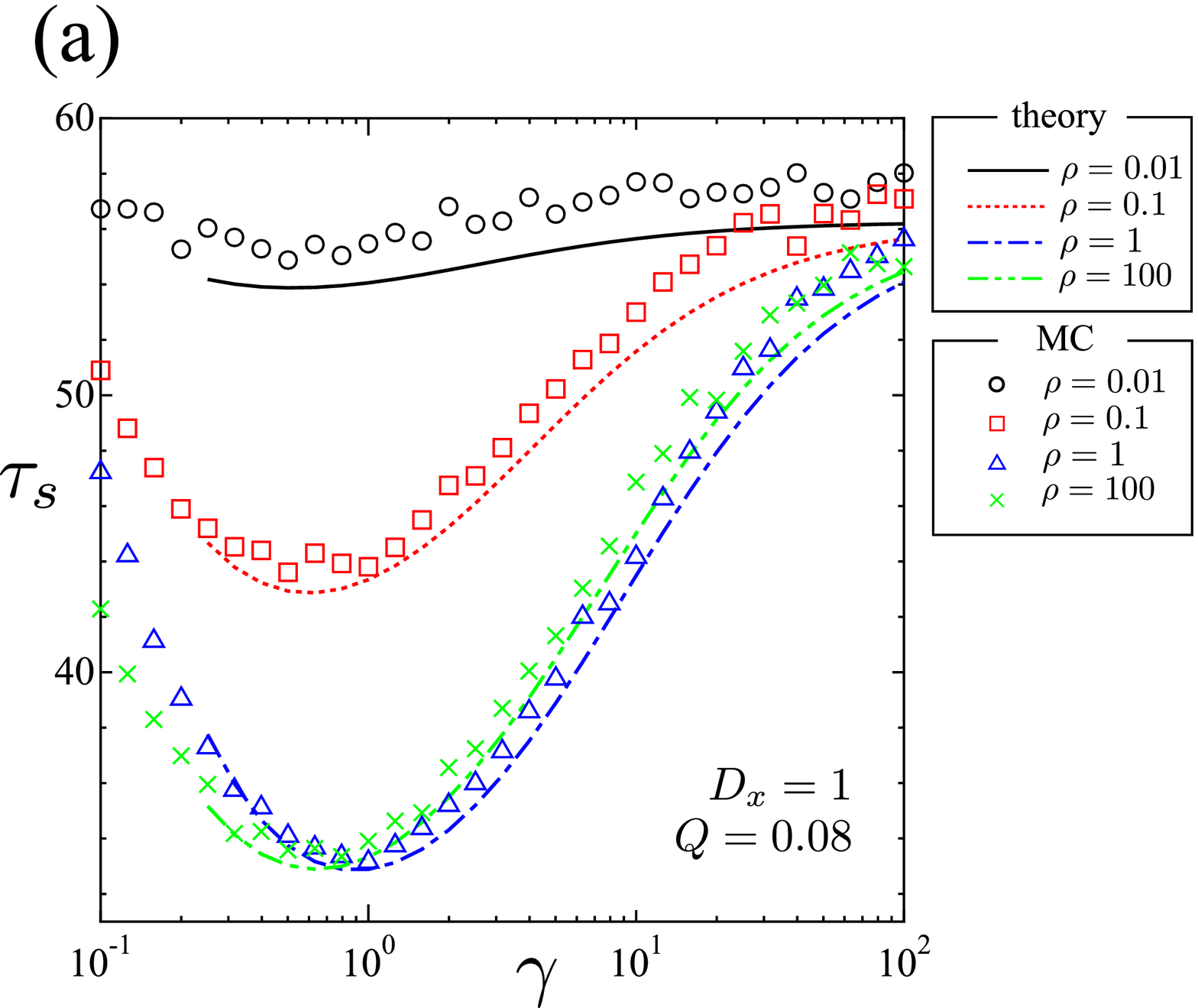}~~~\includegraphics[width=8cm]{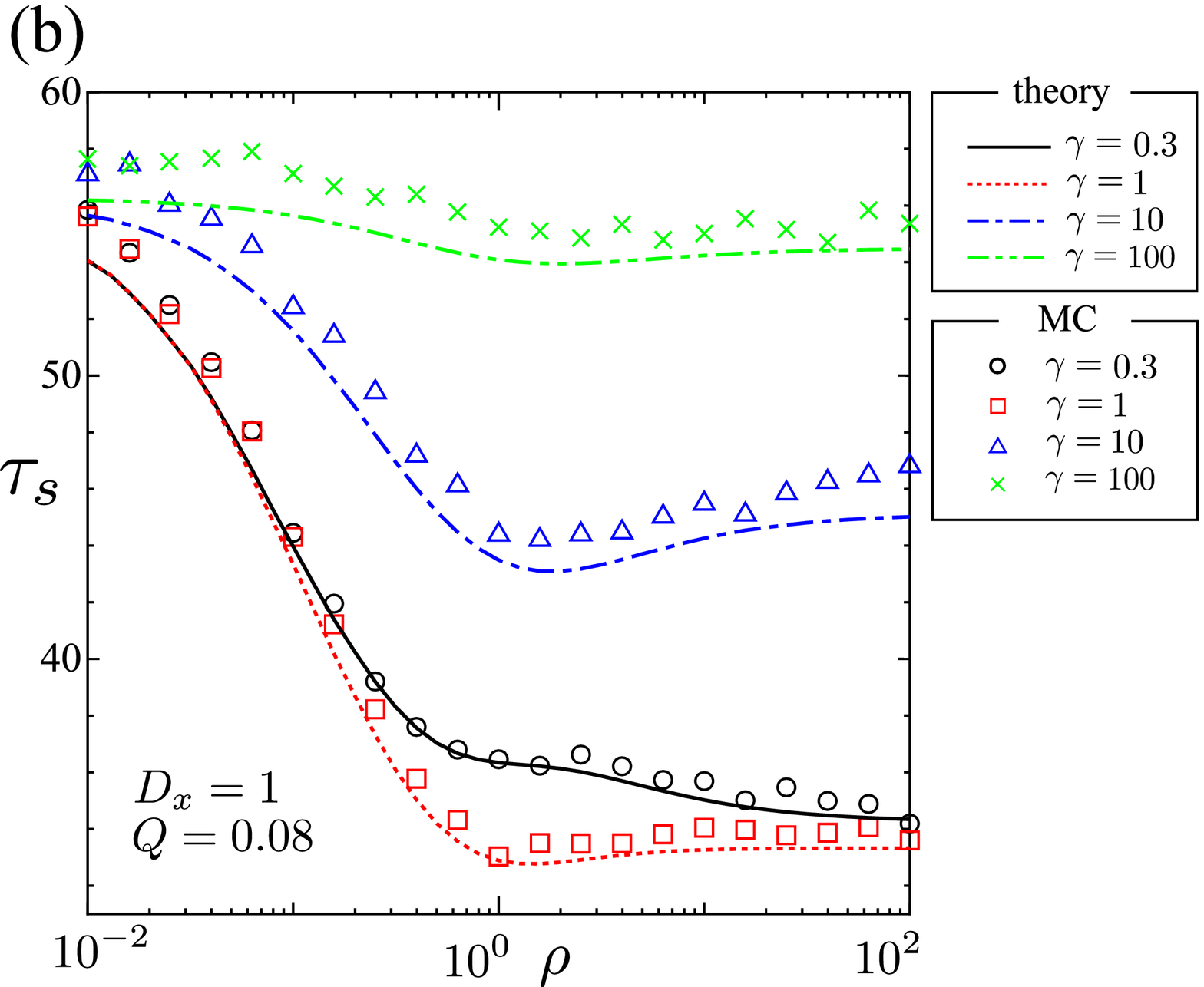} 
\par\end{centering}

\caption{(Color online) MFPT $\tau_{s}$ as a function of (a) the relaxation
rate $\gamma$ and (b) the squared variation coefficient $\rho$.
The lines and symbols denote results of CSE and MC, respectively.
(a) $D_{x}=1$ and $Q=0.08$ with $\rho=0.01$ (solid line and circles),
$0.1$ (dotted line and squares), $1$ (dot-dashed line and triangles),
and $100$ (dot-dot-dashed line and crosses). (b) $D_{x}=1$ and $Q=0.08$
with $\gamma=0.3$ (solid line and circles), $1$ (dotted line and
squares), $10$ (dot-dashed line and triangles), and $100$ (dot-dot-dashed
line and crosses). The MC results are calculated as averages of $20000$
escape events. \label{fig:MFPT_as_gamma}}
\end{figure}

In practical calculation of Eq.~(\ref{eq:linear_eq_eigen}), we increase
$K$ and $L$ until the eigenvalues converge. In addition, we carry
out MC simulations to verify the reliability of the eigenvalue-based
approximation. MFPT of MC is calculated from the average of the first
passage time (FPT) of $20000$ escape events. For sufficiently small
noise intensity, $\tau_{s}$ can be approximated by MFPT $\tau$ from
$-1$ to $0$ because the MFPT dependence on starting points exists
only in a narrow boundary layer around the separatix \cite{Jung:1993:PeriodicSystem}.
In MC calculation, the initial value is $x=-1$, and $s$ has a Gaussian
distribution $\mathcal{N}(\alpha,D_{s})$ with mean $\alpha$ and
variance $D_{s}$. Fig.~\ref{fig:MFPT_as_gamma} shows the MFPT ($\tau_{s}$)
dependence on $\gamma$ and $\rho$; the theoretical results obtained
using CSE are denoted by lines, and the MC results are denoted by
symbols.

Our model includes four parameters: $\gamma$, $\alpha$, $D_{x}$,
and $D_{s}$. In our model calculations, we use $\gamma$, $\rho$,
$Q$, and $D_{x}$ as the given parameters, where $Q$ and $\rho$
are defined by Eqs.~(\ref{eq:Q_def}) and (\ref{eq:rho_def}), respectively.
When these four parameters are given, $\alpha$ and $D_{s}$ are uniquely
determined as $\alpha=\sqrt{Q/\{D_{x}(1+\rho)\}}$ and $D_{s}=\rho Q/\{D_{x}(1+\rho)\}$.
First, we investigate the $\gamma$ dependence of MFPT with $D_{x}=1$,
$Q=0.08$, and various $\rho$ values. Fig.~\ref{fig:MFPT_as_gamma}(a)
shows MFPT as a function of $\gamma$ with four $\rho$ values: $\rho=0.01$
(solid line and circles), $\rho=0.1$ (dotted line and squares), $\rho=1$
(dot-dashed line and triangles), and $\rho=100$ (dot-dot-dashed line
and crosses). From Fig.~\ref{fig:MFPT_as_gamma}(a), $\tau_{s}$
is U-shaped and has a minimum around $\gamma\simeq1$, which can be
accounted for by an RA effect. The conventional RA phenomenon occurs
in a bistable potential subject to white noise, where the potential
fluctuates owing to time-correlated stochastic processes. On the other
hand, the RA observed in Fig.~\ref{fig:MFPT_as_gamma}(a) is induced
by the noise intensity fluctuation. Because MFPT increases with increasing
potential wall height or decreasing noise intensity (or vice versa),
the effect of noise intensity fluctuation on MFPT is similar to that
of potential fluctuation. This correspondence can qualitatively explain
the occurrence of the RA phenomenon in the present model. RA induced
by a noise intensity fluctuation has been reported previously \cite{Iwaniszewski:2008:RAtemperature};
it was realized by the random telegraph process. As expected, the
$\rho=0.01$ case shows a very small RA effect because the noise intensity
fluctuation is very weak in this case. For larger $\rho$, the RA
effect is larger because the noise intensity fluctuation increases
with $\rho$ {[}Fig.~\ref{fig:path_rho}{]}. In contrast, the RA
effects of $\rho=1$ and $\rho=100$ are nearly similar. Remarkably,
the effect of RA for $\rho=100$ is not larger than that for $\rho=1$,
even though the noise intensity fluctuation is stronger for $\rho=100$
(Fig.~\ref{fig:path_rho}(c) and (d)).

Next, we calculate the $\rho$ dependence of MFPT by varying $\rho$
while keeping the effective intensity $Q$ constant. Fig.~\ref{fig:MFPT_as_gamma}(b)
shows MFPT as a function of $\rho$ with four $\gamma$ values: $\gamma=0.3$
(solid line and circles), $\gamma=1$ (dotted line and squares), $\gamma=10$
(dot-dashed line and triangles), and $\gamma=100$ (dot-dot-dashed
line and crosses). For $\gamma=0.3$, $\tau_{s}$ decreases as a function
of $\rho$. On the other hand, $\tau_{s}$ has a minimum around $\rho\sim1$
for $\gamma=1$, $10$, and $100$ (the depth of the minimum is smaller
for larger $\gamma$). As explained, the RA phenomenon is referred
to as the existence of the minimum as a function of the relaxation
rate. The strength of the RA effect can be measured by the magnitude
of the minima. In all $\rho$ cases in Fig.~\ref{fig:MFPT_as_gamma}(a),
MFPT is minimum around $\gamma\simeq1$. Therefore, MFPT in Fig.~\ref{fig:MFPT_as_gamma}(b)
with $\gamma=1$ (dotted line) can be identified as the strength of
the RA effect as a function of $\rho$. This indicates that the strength
of the RA effect increases with $\rho$, up to $\rho\simeq1$. A further
increase in $\rho$ does not increase the strength of the RA effect.

In Fig.~\ref{fig:MFPT_as_gamma}, the theoretical results obtained
using CSE (lines) are in agreement with MC simulations (symbols) for
all cases; this verifies the reliability of the approximation scheme.

\section{Stochastic Resonance\label{sub:SR}}

Next, we study SR \cite{Benzi:1981:SR,McNamara:1989:SR,Jung:1990:MFCforPeriodic,Jung:1991:AmpSR,Gammaitoni:1998:SR,Jia:2001:SRwithAddMult,Mantegna:2000:LinNlinSR,McDonnell:2008:SRBook,McDonnell:2009:SR,Agudov:2010:MonostableSR}
in our model. SR is an intriguing phenomenon, and it plays an important
role in systems accompanied by noise; hence, it has been studied extensively
in various configurations. In particular, biological applications
of SR have attracted considerable attention, and they have been confirmed
experimentally and theoretically \cite{Longtin:1991:SRinNeuron,Hanggi:2002:SRinBio,Priplata:2002:SRinBalance,Priplata:2003:ValanceControl}
because biological mechanisms occur in noisy environments. We calculate
the spectral amplification factor of SR with a periodic input under
additive SIN. Specifically, we employ linear response approximation
\cite{Dykman:1993:LRinSR} to calculate the quantity. For a sufficiently
small driving force, linear response approximation can be used to
investigate SR.

We assume that the system of interest is modulated by an external
input $\varepsilon\exp(-\mathrm{i}\Omega t)$, where $\varepsilon$
and $\Omega$ are the input strength and the angular frequency, respectively.
A Langevin equation is given by 
\begin{equation}
\frac{dx}{dt}=f(x)+\varepsilon\exp(-\mathrm{i}\Omega t)+s\xi_{x}(t),\label{eq:periodic_Langevin}
\end{equation}
 and Eq.~(\ref{eq:SIN_Langevin}), where $f(x)=x-x^{3}$. The FPE
of Eqs.~(\ref{eq:periodic_Langevin}) and (\ref{eq:SIN_Langevin})
is 
\begin{equation}
\frac{\partial}{\partial t}P(x,s;t)=\mathscr{L}_{p}P(x,s;t),\label{eq:FPE_periodic}
\end{equation}
 with 
\begin{equation}
\mathscr{L}_{p}=\mathscr{L}_{0}+\mathscr{L}_{1}\varepsilon\exp(-\mathrm{i}\Omega t),\hspace{1em}\mathscr{L}_{1}=-\frac{\partial}{\partial x},\label{eq:FPop_input}
\end{equation}
 where $\mathscr{L}_{0}$ is defined in Eq.~(\ref{eq:FPE_add}).
We assume that $\varepsilon$ is sufficiently small for the system
to be well approximated by the linear response. Let $P_{as}(x,s;t)$
be an asymptotic solution ($t\rightarrow\infty$) of Eq.~(\ref{eq:FPE_periodic}).
According to the Floquet theorem, $P_{as}(x,s;t)$ is a periodic function
having the same period as the input: 
\begin{equation}
P_{as}(x,s;t)=P_{as}(x,s;t+T),\label{eq:Pas_periodic}
\end{equation}
 where $T$ is the period {[}$T=2\pi/\Omega${]}. According to Eq.~(\ref{eq:Pas_periodic})
and the linear response approximation, we can expand $P_{as}(x,s;t)$
as 
\begin{equation}
P_{as}(x,s;t)=P_{0}(x,s)+P_{1}(x,s)\varepsilon\exp(-\mathrm{i}\Omega t).\label{eq:LR_expansion}
\end{equation}
 From a normalization condition, $P_{1}(x,s)$ must satisfy 
\begin{equation}
\int dx\int ds\, P_{1}(x,s)=0.\label{eq:P1_normalization}
\end{equation}
 Substituting Eq.~(\ref{eq:LR_expansion}) into Eq.~(\ref{eq:FPE_periodic})
and comparing the order of $\varepsilon$, we obtain the following
coupled equations: 
\begin{eqnarray}
O(1) &  & \mathscr{L}_{0}P_{0}(x,s)=0,\label{eq:LR_order0}\\
O(\varepsilon) &  & \mathscr{L}_{0}P_{1}(x,s)+\mathscr{L}_{1}P_{0}(x,s)=-\mathrm{i}\Omega P_{1}(x,s).\label{eq:LR_order1}
\end{eqnarray}
 Eq.~(\ref{eq:LR_order0}) is identical to the equation for stationary
distributions {[}Eq.~(\ref{eq:stat_FPE_op}){]}. Following the procedure
for stationary distributions (Sec.~\ref{sec:stat_dist}), we expand
$P_{1}(x,s)$ in terms of the orthonormal complete set. Using the
relation $P_{as}(x,s;t)=P_{as}(-x,s;t+T/2)$ in Eq.~(\ref{eq:FPop_input}),
$P_{1}(x,s)$ admits the odd symmetry expansion: 
\begin{equation}
P_{1}(x,s)=\varphi_{0}(x)\psi_{0}(s)\sum_{k=0}^{K}\sum_{\ell=0}^{L}G_{k,\ell}\varphi_{2k+1}(x)\psi_{\ell}(s),\label{eq:P1_expansion}
\end{equation}
 where $G_{k,\ell}$ are coefficients. Note that Eq. (\ref{eq:P1_expansion})
automatically satisfies Eq.~(\ref{eq:P1_normalization}) because
of the orthonormality. Following the same procedures as those in Secs.~\ref{sec:stat_dist}
and \ref{sec:MFPT}, Eqs.~(\ref{eq:LR_order0}) and (\ref{eq:LR_order1})
can be represented as the following linear algebraic equation in terms
of $G_{k,\ell}$: 
\begin{eqnarray}
0 & = & \zeta\sqrt{2(2k+1)}C_{k,\ell}+G_{k,\ell}\left\{ 2k+1-\frac{3}{2\zeta^{2}}(2k+1)^{2}-\gamma\ell+\mathrm{i}\Omega\right\} \nonumber \\
 &  & +G_{k-1,\ell}\sqrt{2k(2k+1)}\left[1-\frac{3k}{\zeta^{2}}+2\zeta^{2}D_{x}\left\{ \alpha^{2}+2D_{s}\left(\ell+\frac{1}{2}\right)\right\} \right]\nonumber \\
 &  & -G_{k+1,\ell}\frac{2k+1}{2\zeta^{2}}\sqrt{(2k+2)(2k+3)}-G_{k-2,\ell}\frac{1}{2\zeta^{2}}\sqrt{(2k+1)2k(2k-1)(2k-2)}\nonumber \\
 &  & +2G_{k-1,\ell+2}\zeta^{2}D_{x}D_{s}\sqrt{2k(2k+1)(\ell+1)(\ell+2)}+2G_{k-1,\ell-2}\zeta^{2}D_{x}D_{s}\sqrt{2k(2k+1)\ell(\ell-1)}\nonumber \\
 &  & +4G_{k-1,\ell-1}\zeta^{2}D_{x}\alpha\sqrt{2D_{s}k(2k+1)\ell}+4G_{k-1,\ell+1}\zeta^{2}D_{x}\alpha\sqrt{2D_{s}k(2k+1)(\ell+1)}.\label{eq:P1_linear}
\end{eqnarray}
 $C_{k,\ell}$ has already been calculated in Eq.~(\ref{eq:P0_linear})
for the stationary distributions.

From Eq.~(\ref{eq:LR_expansion}), the time-dependent asymptotic
average of $x$ is given by 
\begin{eqnarray}
\left\langle x(t)\right\rangle _{as} & = & \int dx\int ds\, xP_{as}(x,s;t),\nonumber \\
 & = & \left\langle x\right\rangle _{0}+\left\langle x\right\rangle _{1}\varepsilon\exp(-\mathrm{i}\Omega t),\label{eq:x_as}
\end{eqnarray}
 with 
\[
\left\langle x\right\rangle _{0}=\int dx\int ds\, xP_{0}(x,s),\hspace{1em}\left\langle x\right\rangle _{1}=\int dx\int ds\, xP_{1}(x,s),
\]
 where $\left\langle x\right\rangle _{0}=0$ owing to the symmetry.
The susceptibility $\chi$ is defined as the proportional coefficient
of the input signal, which is given by $\chi=\left\langle x\right\rangle _{1}.$
There are several approaches to calculating the susceptibility, e.g.,
the fluctuation-dissipation relation \cite{Jung:1993:PeriodicSystem}
or the moment method \cite{Evstigneev:2001:MM4Periodic,Kang:2003:DuffSRMoment}.
Using the orthonormal and complete relations, the susceptibility is
\begin{equation}
\chi=\frac{G_{0,0}}{\sqrt{2}\zeta}.\label{eq:susceptibility}
\end{equation}
 Let us consider a cosinusoidal input $\varepsilon\cos(\Omega t)$.
$\left\langle x(t)\right\rangle _{as}$ for this case is 
\begin{equation}
\left\langle x(t)\right\rangle _{as}=\left\langle x\right\rangle _{0}+|\chi|\varepsilon\cos(\Omega t+\theta),\hspace{1em}\theta=-\arctan\left(\frac{\mathrm{Im}(\chi)}{\mathrm{Re}(\chi)}\right),\label{eq:empirical_average_path}
\end{equation}
 where $\theta$ is the phase. We evaluate the spectral amplification
as $|\chi|^{2}=|G_{0,0}|^{2}/(2\zeta^{2})$.

\begin{figure}
\begin{centering}
\includegraphics[width=7cm]{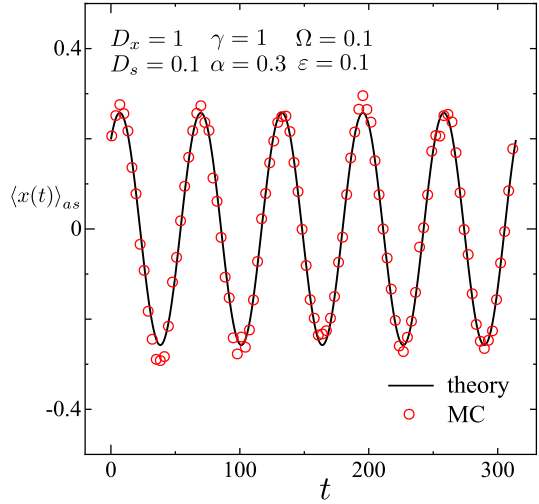} 
\par\end{centering}

\caption{(Color online) $\left\langle x(t)\right\rangle _{as}$ of CSE results
{[}Eqs.~(\ref{eq:susceptibility}) and (\ref{eq:empirical_average_path}){]}
(solid line) and MC simulations as the average of $2000$ trajectories
(circles). The parameters are $D_{x}=1$, $D_{s}=0.1$, $\gamma=1$,
$\alpha=0.3$, and $\Omega=0.1$, and $\varepsilon=0.1$ for MC. \label{fig:average_path}}
\end{figure}

We perform MC simulations to verify the reliability of the linear
response approximation. For MC simulations, a method in Ref.~\cite{Li:2009:MultiplicativeSR}
was employed. The averages of $2000$ trajectories were calculated
and the susceptibility was estimated by their variance {[}Eq.~(\ref{eq:empirical_average_path}){]}
(the method of moments estimation). Fig.~\ref{fig:average_path}
shows $\left\langle x(t)\right\rangle _{as}$ calculated by Eqs.~(\ref{eq:susceptibility})
and (\ref{eq:empirical_average_path}) (solid line) and MC simulations
(circles). We observe excellent agreement between them, which verifies
the reliability of the linear response approximation.

\begin{figure}
\begin{centering}
\includegraphics[width=8cm]{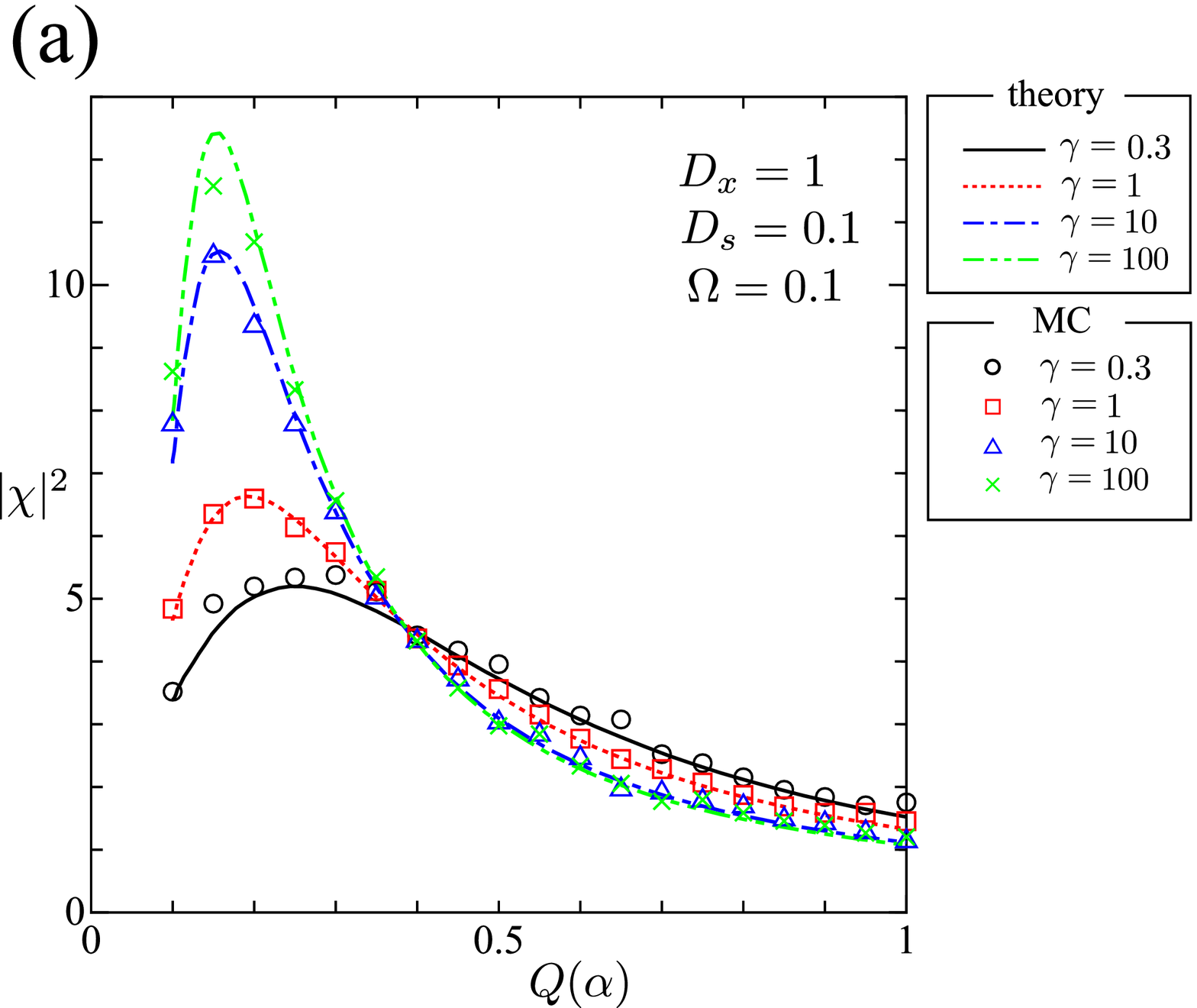}~~~\includegraphics[width=8cm]{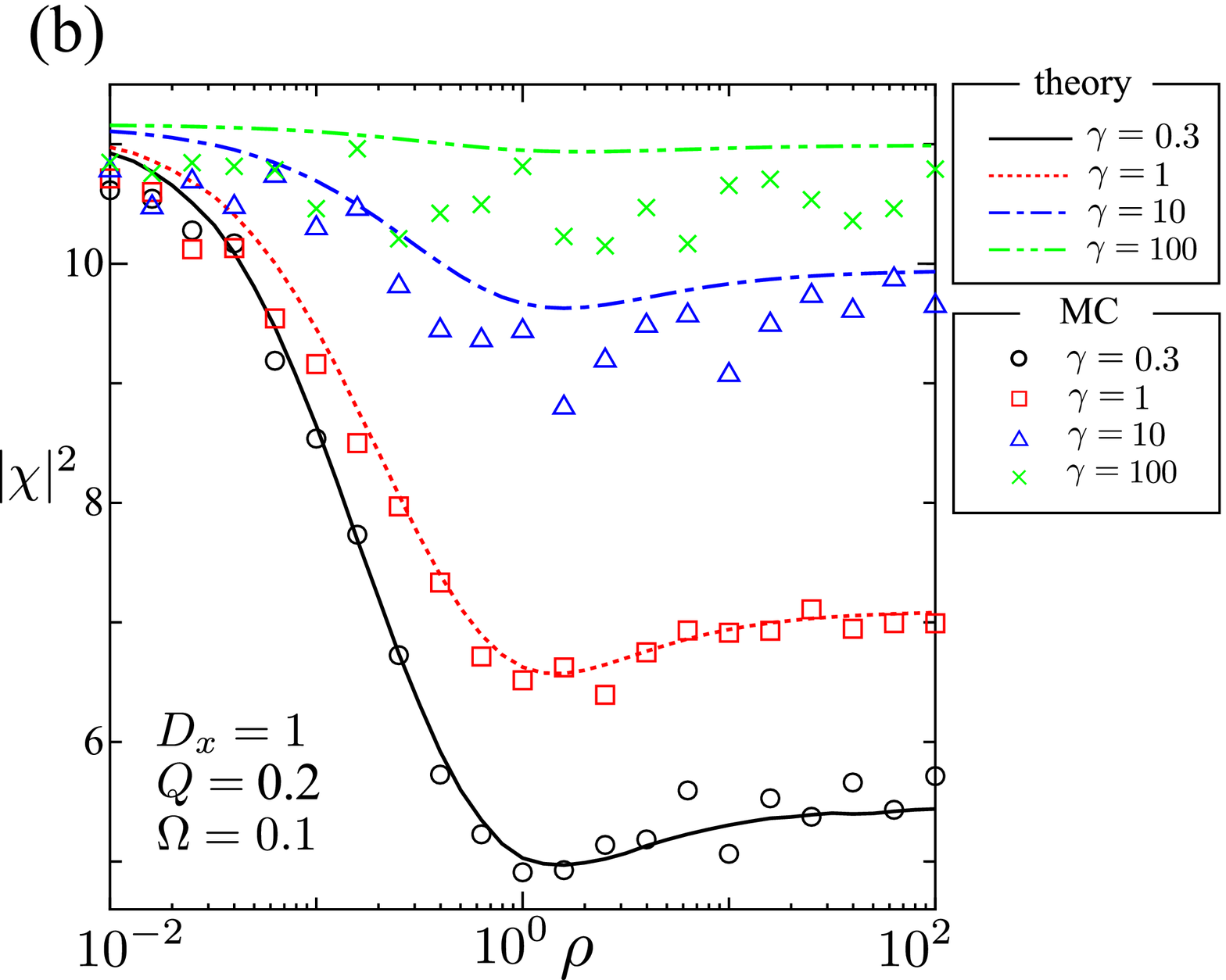} 
\par\end{centering}

\begin{centering}
\includegraphics[width=8cm]{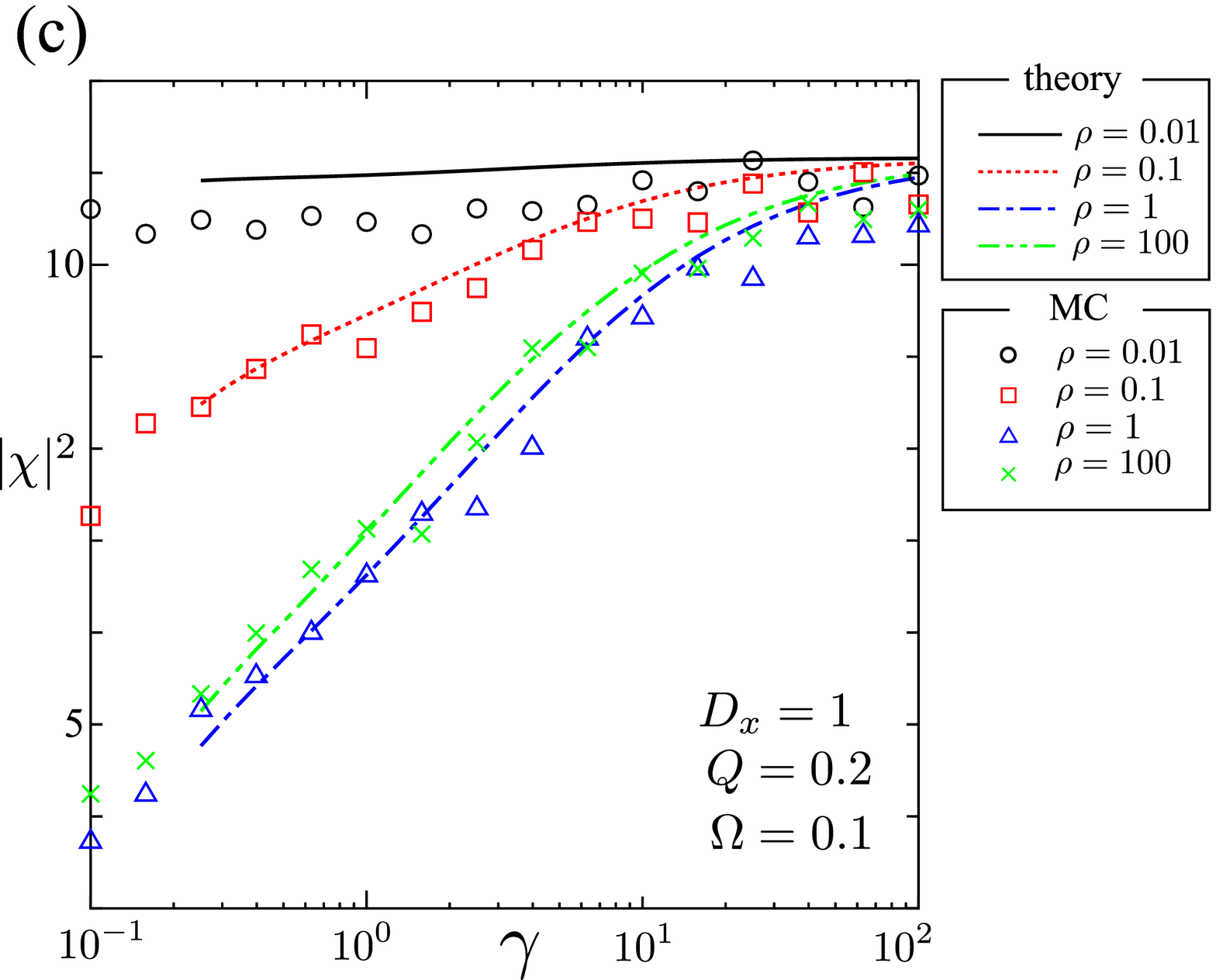} 
\par\end{centering}

\caption{(Color online) (a) Spectral amplification factor $|\chi|^{2}$ as
a function of (a) the effective noise intensity $Q(\alpha)$ (vary
$\alpha$ while keeping $D_{s}$ constant), (b) the square variation
coefficient $\rho$, and (c) the relaxation rate $\gamma$. The lines
and symbols denote CSE and MC results, respectively. (a) $D_{x}=1$,
$D_{s}=0.1$ and $\Omega=0.1$ with $\gamma=0.3$ (solid line and
circles), $1$ (dotted line and squares), $10$ (dot-dashed line and
triangles), and $100$ (dot-dot-dashed line and crosses). (b) $D_{x}=1$,
$Q=0.2$ and $\Omega=0.1$ with $\gamma=0.3$ (solid line and circles),
$1$ (dotted line and squares), $10$ (dot-dashed line and triangles),
and $100$ (dot-dot-dashed line and crosses). (c) $D_{x}=1$, $Q=0.2$,
and $\Omega=0.1$ with $\rho=0.01$ (solid line and circles), $0.1$
(dotted line and squares), $1$ (dot-dashed line and triangles), and
$100$ (dot-dot-dashed line and crosses). MC results with $\varepsilon=0.1$
are calculated using the variance of the average of $2000$ trajectories
(the method of moments estimation). \label{fig:SR_amp_Dx}}
\end{figure}

Fig.~\ref{fig:SR_amp_Dx} shows the dependence of the spectral amplification
factor $|\chi|^{2}$ on $Q$, $\rho$, and $\gamma$, where theoretical
results obtained using CSE are denoted by lines and MC results are
denoted by symbols. The MC results were in good agreement with those
of CSE, verifying their reliability.

Specifically, Fig.~\ref{fig:SR_amp_Dx}(a) shows the $|\chi|^{2}$
dependence on $Q(\alpha)=D_{x}(D_{s}+\alpha^{2})$ ($\alpha$ is varied
while keeping $D_{x}$ and $D_{s}$ constant) with four $\gamma$
values: $\gamma=0.3$ (solid line and circles), $\gamma=1$ (dotted
line and squares), $\gamma=10$ (dot-dashed line and triangles), and
$\gamma=100$ (dot-dot-dashed line and crosses) with $D_{x}=1$, $D_{s}=0.1$,
and $\Omega=0.1$. Here, $|\chi|^{2}$ achieves a maximum around $Q(\alpha)=0.2$,
and the maximum is larger for larger $\gamma$. SIN approaches white
noise for $\gamma\rightarrow\infty$, indicating that the strength
of the SR effect is maximized under white noise. On the other hand,
$|\chi|^{2}$ in the range $Q(\alpha)\gtrsim0.4$ has a different
tendency, i.e., $|\chi|^{2}$ is larger for smaller $\gamma$. Although
the peaks of $|\chi|^{2}$ at $Q(\alpha)\simeq0.2$ are smaller for
smaller $\gamma$, SIN can induce better performance when the noise
intensity exceeds $Q(\alpha)\simeq0.4$.

Next, we calculate $|\chi|^{2}$ as a function of $\rho$ with four
$\gamma$ values: $\gamma=0.3$ (solid line and circles), $\gamma=1$
(dotted line and squares), $\gamma=10$ (dot-dashed line and triangles)
and $\gamma=100$ (dot-dot-dashed line and crosses). We vary $\rho$
while keeping the effective noise intensity $Q$ constant. Because
the spectral amplification factor $|\chi|^{2}$ is maximum as a function
of the effective noise intensity in SR, its strength can be measured
by the magnitude of the maxima. The maxima in Fig.~\ref{fig:SR_amp_Dx}(a)
are located around $Q\simeq0.2$; hence, we fixed $Q=0.2$ and investigated
$|\chi|^{2}$ dependence on $\rho$ in Fig.~\ref{fig:SR_amp_Dx}(b)
($D_{x}$ and $\Omega$ are the same as as those in Fig.~\ref{fig:SR_amp_Dx}(a)).
Accordingly, $|\chi|^{2}$ of Fig.~\ref{fig:SR_amp_Dx}(b) can be
identified as the strength of the SR effect as a function of $\rho$.
Because SIN reduces to white noise as $\gamma\rightarrow\infty$,
$|\chi|^{2}$ as a function of $\rho$ does not change for $\gamma=100$.
On the other hand, $|\chi|^{2}$ is more strongly affected by $\rho$
for smaller $\gamma$. SIN also reduces to white noise as $\rho\rightarrow0$,
and $|\chi|^{2}$ increases as $\rho\rightarrow0$ in all cases. We
observe non-monotonic behavior of $|\chi|^{2}$ as a function of $\rho$,
i.e., the strength of the SR effect is minimized around $\rho\simeq1$.
Remarkably, the effect of the input signal is minimized around $\rho\simeq1$,
even though the strength of the noise intensity fluctuation is monotonic
as a function of $\rho$.

Fig.~\ref{fig:SR_amp_Dx}(c) shows $|\chi|^{2}$ as a function of
$\gamma$ for four $\rho$ values: $\rho=0.01$ (solid line and circles),
$\rho=0.1$ (dotted-line and squares), $\rho=1$ (dot-dashed line
and triangles), and $\rho=100$ (dot-dot-dashed line and crosses)
with $D_{x}=1$, $Q=0.2$, and $\Omega=0.1$. In all cases, $|\chi|^{2}$
increases as a function of $\gamma$; therefore, the SR effect achieves
a maximum under white noise.

\section{Discussion\label{sec:discussion}}

\begin{figure}
\begin{centering}
\includegraphics[width=7cm]{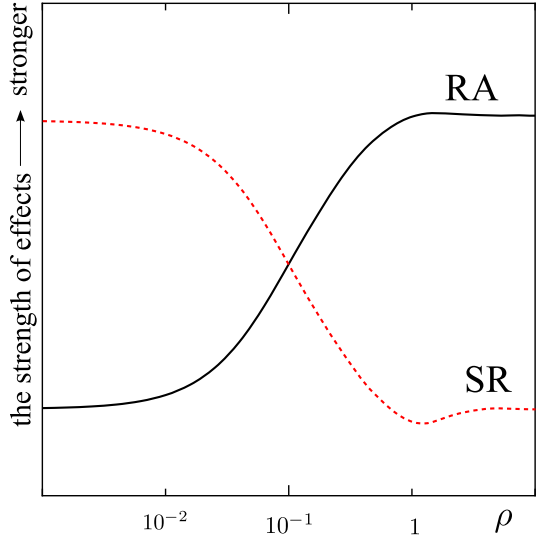} 
\par\end{centering}

\caption{(Color online) Illustrative description of the strength of the RA
and SR effects as a function of the squared variation coefficient
$\rho$. We observe a trade-off relation between the strength of RA
and SR. \label{fig:relation_SR_RA}}
\end{figure}

In Secs.~\ref{sec:MFPT} and \ref{sub:SR}, we have shown that the
strength of the RA and SR effects exhibits non-monotonic behavior
as a function of the squared variation coefficient $\rho$. Furthermore,
the strength of RA and SR effects is enhanced in different $\rho$
regions. The strength of the RA effect is maximum around $\rho\gtrsim1$,
whereas that of the SR effect is stronger for $\rho\lesssim10^{-2}$.
On the other hand, the strength of the SR and RA effects is very weak
in regions of $\rho\simeq1$ and $\rho\lesssim10^{-2}$, respectively.
These results shows that strength of these two effects has a trade-off
relation in terms of $\rho$. An illustrative description of the trade-off
relation between the strength of RA and SR effects is shown in Fig.~\ref{fig:relation_SR_RA},
where the solid and dotted lines represent the strength of the RA
and SR effects, respectively, as a function of $\rho$.

Langevin equations have been extensively applied to stochastic biochemical
reactions such as gene expression \cite{Koern:2005:GeneNoiseReview}
and neuronal response. In a zeroth-order approximation, these biological
mechanisms can be modeled using a bistable potential \cite{Wilhelm:2009:Bistability}.
Biological mechanisms are subject to many fluctuations having different
time-scales. It has been reported theoretically and experimentally
that RA and SR are expected to play important roles in biological
mechanisms. RA can minimize the delays in signal detection, which
improves the response to signals. On the other hand, SR is responsible
for accurate signal detection in noisy environments. These two factors
are important in signal transmission, and our results indicate that
their importance can be tuned with $\rho$. The results presented
above may provide us with a new insight into the analyses of stochastic
aspects of biological mechanisms.

\section{Concluding Remarks\label{sec:remarks}}

In the present paper, we employed CSE to calculate stationary distributions,
MFPT, and the spectral amplification factor. In our previous study
\cite{Hasegawa:2010:SIN}, we used adiabatic elimination to derive
a time evolution equation. CSE is advantageous in that the ranges
of the relaxation-rate $\gamma$ and the noise intensity are not limited,
as opposed to the adiabatic elimination-based method, which is valid
for $\gamma\gg1$. In addition, CSE enables us to calculate quantities
such as MFPT and the spectral amplification factor. On the other hand,
using adiabatic elimination, we can calculate stationary distributions
in the closed form, and it can be used for general non-linear drift
terms. In contrast, CSE can only handle polynomial drift terms, for
which stationary distributions are obtained by a numerical method.
Both approaches are complementary. From the MFPT calculation, we identified
the RA phenomenon as a function of $\gamma$. We also showed that
the strength of the RA effect is highly dependent on the squared variation
coefficient $\rho$, and that the strength of the SR effect as a function
of $\rho$ is minimum around $\rho\simeq1$. These results indicate
that $\rho$, the ratio between the variance and mean of the noise
intensity modulating process {[}Eq.~(\ref{eq:rho_def}){]}, has a
crucial impact on the RA and SR effects.

Because CSE can be used for polynomial drift terms with arbitrary
magnitudes of relaxation rate and noise intensity, the analysis described
in this paper can be applied to various real-world phenomena. Furthermore,
we focused on periodic SR, in which the system of interest is modulated
by a periodic input. With regard to biological cases, the investigation
of aperiodic SR \cite{Collins:1995:ASRinExcite,Collins:1996:AperiodicSR}
is important. We plan to investigate this subject in the future.

\section*{Acknowledgments}

This work was supported by a Grand-in-Aid for Scientific Research
on Priority Areas (17017006) and a Grant-in-Aid for Young Scientists
B (23700263).

\appendix

\section{Correlation function}

Here, we calculate the correlation function of SIN. By definition,
the correlation function is given by 
\begin{equation}
\left\langle s(t)\xi_{x}(t)s(t^{\prime})\xi_{x}(t^{\prime})\right\rangle =\int dsds^{\prime}d\xi_{x}d\xi_{x}^{\prime}\,\left[ss^{\prime}\xi_{x}\xi_{x}^{\prime}P(s,\xi_{x};t|s^{\prime},\xi_{x}^{\prime};t^{\prime})P(s^{\prime},\xi_{x}^{\prime};t^{\prime})\right].\label{eq:cor_def}
\end{equation}
 Since $s(t)$ and $\xi_{x}(t)$ are independent, Eq.~(\ref{eq:cor_def})
becomes 
\begin{eqnarray}
\left\langle s(t)\xi_{x}(t)s(t^{\prime})\xi_{x}(t^{\prime})\right\rangle  & = & \int dsds^{\prime}d\xi_{x}d\xi_{x}^{\prime}\,\left[ss^{\prime}\xi_{x}\xi_{x}^{\prime}P(s;t|s^{\prime};t^{\prime})P(s^{\prime};t^{\prime})P(\xi_{x};t|\xi_{x}^{\prime};t^{\prime})P(\xi_{x}^{\prime};t^{\prime})\right],\nonumber \\
 & = & \left\langle s(t)s(t^{\prime})\right\rangle \left\langle \xi_{x}(t)\xi_{x}(t^{\prime})\right\rangle ,\label{eq:cor_SIN1}
\end{eqnarray}
 where the correlation function of $s(t)$ is calculated as 
\begin{equation}
\left\langle s(t)s(t^{\prime})\right\rangle =D_{s}\exp\left(-\gamma|t-t^{\prime}|\right)+\alpha^{2}.\label{eq:cor_st}
\end{equation}
 From Eqs.~(\ref{eq:cor_SIN1}) and (\ref{eq:cor_st}), we obtain
\begin{eqnarray}
\left\langle s(t)\xi_{x}(t)s(t^{\prime})\xi_{x}(t^{\prime})\right\rangle  & = & 2D_{x}\left\{ D_{s}\exp\left(-\gamma|t-t^{\prime}|\right)+\alpha^{2}\right\} \delta(t-t^{\prime}),\nonumber \\
 & = & 2D_{x}(D_{s}+\alpha^{2})\delta(t-t^{\prime}),\\
 & = & 2Q\delta(t-t^{\prime}),\label{eq:cor_SIN2}
\end{eqnarray}
 where $Q$ is the effective intensity defined by Eq.~(\ref{eq:Q_def}).
From Eq.~(\ref{eq:cor_SIN2}), the intensity of SIN is in agreement
with the effective intensity $Q$, which is calculated via adiabatic
elimination \cite{Hasegawa:2010:SIN}.

\end{document}